\documentclass[twocolumn,a4paper,10pt]{article}
\usepackage{amsmath}
\usepackage{amsthm}
\usepackage{enumerate}
\title{\textbf{An investigation of Lorentz transformation}}
\author{Farid Sh\"ahandeh\\
        \textit{Physics Department, Faculty of Science, I. K. I. University}\\
        \textit{Imam Khomeini Blvd., Qazvin, Iran.}\\
        \small E-mail address: \ttfamily shahandeh@ikiu.ac.ir
        }
\date{}
\begin{document}
\twocolumn[
 \begin{@twocolumnfalse}
  \maketitle
  \begin{abstract}
A new method of derivation of Lorentz Transformation (LT) is given based on both axioms of special relativity (SR) and physical intuitions. The essence of the transformation is established and the crucial role played by the presumptions is presented for clarification. I consider the most general form of transformations between two sets of events in two inertial reference frames and use the most basic properties expected from such a transformation together with the principle of the constancy of the velocity of light to derive LT. The method is very simple, succinct and useful for students trying a better understanding of the subject.
  \end{abstract}
 \end{@twocolumnfalse}
]
\section{Introduction}
Because of its very peculiar consequences, at the time of presenting SR, many physicists those who had advocated for Newtonian physics tried to expose a logical flaw in the theory. Today there are many empirical evidences for these conclusions, however, the debates continue.\cite{TS}

Many authors argued the way in which Einstein derived LT from two basic postulates for flaws in it.\cite{TS,VS} On the other hand, others stated different derivations to that of Einstein\cite{WR} escaping the ambiguities. Regardless of the mathematical procedures they used, the underlying assumptions of the theory, somehow, lost their clarity providing different misunderstandings of the subject and its consequences.\cite{NH}

In this paper, I present a review of the subject (see \cite{AJP1,AJP2,AJP3,AJP4}) using a demystified mathematical tool, clarifying the use of physical intuition and the principles of SR, leading to a simple derivation of LT. The method I use here helps further discussions for students interested in the subject. 
\section{Axioms of SR}
In his 1905 paper, Einstein proposed two postulates:\cite{AE}
\begin{enumerate}[I.]
  \item The laws of physics are the same to all inertial observers.
  \item The speed of light is the same to all inertial observers.\cite{HG}
\end{enumerate}

Some authors discussed the necessity of the axioms \cite{OD} and some others reduced them to just one extending the so-called synchronization process~\cite{Sf}, however, I will show that introducing the speed of light as a universal constant is unavoidable. Instead of discussing the axioms, I focus here on the use of physical intuitions emphasizing on the role they play behind any derivation of LT. From now on I refer to an inertial frame of reference as a `reference frame' or just a `frame.'
\section{Events and frames of reference}
To find out the way in which different inertial frames of reference are related to each other, we define individual thought objects called `events.' Events are the core of the theory as they can be distinguished by `observers' in all frames. However, the term observer is not in the common sense a person `seeing' the things. An observer is defined to be a set of standard `rods' and `watches' or any set of standard apparatuses for measuring length and time (c.~f.~Ref.~3.)

In each reference frame, we refer to any single event in that frame using a multiple of numbers called the `coordinates' of the event. The coordinate of an event, specifies its position and time of occurrence with respect to that frame.

Considering the simple one dimensional case, for example, we may identify an individual event in the frame one by the pair $\left(x,y\right)_{1}$ as its coordinates in that frame.

If an event occurs in frame one with coordinates $\left(x,y\right)_{1}$, it corresponds to the event $\left(x,y\right)_{2}$ in frame two. This assumption might be called the `correspondence' hypothesis. That is to say:
\begin{quote}
Corresponding to any set of events in one specific reference frame, there exists a set of equivalent events in all other frames.
\end{quote}
In other words, there exists a one-to-one mapping from the set of events $\left\{\left(x_{i},y_{i}\right)\right\}^{N}_{i=0}$ in frame one to any inertial reference frame which is moving with the relative speed $v$. In the rest of this paper, our goal is to find this mapping and investigating whether it is unique or not.

From homogeneity of space it is reasonable to consider the direction of the relative motion along the common x-axis, thus using the pair $\left(x_{i},y_{i}\right)_{j}$ suffices for our current purpose. In this way, we can describe the most general mapping between frames by a two-by-two matrix which is a linear transformation:\cite{DW}
\begin{equation} \label{DEF}
  \begin{bmatrix}
         x_i \\
         ct_i
  \end{bmatrix}_{k}=T\left(v_{kj}\right)%
  \begin{bmatrix}
         x_i \\
         ct_i
  \end{bmatrix}_{j}
\end{equation}
(\emph{no summation over indexes}), where the transformation matrix is defined to be:
\begin{equation} \label{T}
 T\left(v_{kj}\right)=
  \begin{bmatrix}
         T_{11}\left(v_{kj}\right) & T_{12}\left(v_{kj}\right) \\
         T_{21}\left(v_{kj}\right) & T_{22}\left(v_{kj}\right)
  \end{bmatrix}
\end{equation}
In this notation, the frame $k$ is moving with the speed $v_{kj}$ with respect to the frame $j$ in the positive direction of the common $x$-axis. Equivalently, the frame $j$ moving with respect to the frame $k$ with the speed $v_{jk}=-v_{kj}$. From now on, we use the shorthands $v$ for $v_{kj}$ and $-v$ for $v_{jk}$. Moreover, the constant $c$ is just a matter of dimensional correction. This coefficient, of course, must have the dimension of velocity and needs to be independent of the frames. Thus, by the second axiom, the speed of light is a proper choice.
\section{Intuitions}
In this section we state the basic intuitive properties expected from the mapping between sets of events among reference frames:
\begin{enumerate}
	\item In the limit of $v\rightarrow0$ this mapping must correspond to unity, or just:
	\begin{equation}
	 lim_{v\rightarrow0}T\left(v\right)=1
	\end{equation}
This property asserts that when two frames have no relative motion, the set of events in the former frame must be equivalent to the same set of events in the latter frame, as in everyday experience.
	\item 	By successive use of mappings corresponding to $v$ and $-v$, one must achieve the unity transformation:
	\begin{equation}
	 T\left(v\right)T\left(-v\right)=T\left(-v\right)T\left(v\right)=1
	\end{equation}
which is equivalent to:
	\begin{equation} \label{INV}
	 T^{-1}\left(v\right)=T\left(-v\right)
	\end{equation}
This is also obvious, because switching from one frame to another is equivalent to reversing the direction of relative motion. Also, switching twice between two events must lead to the original one.
	\item Successive events which are only time-part separated, must correspond to a measure of relative speed $v$ in the other frames:
	\begin{eqnarray}
	 T\left(v\right)\left(
	 \begin{bmatrix}
         x_{i} \\
         ct_{i+1}
   \end{bmatrix}_{j}-
   \begin{bmatrix}
          x_i \\
          ct_i
   \end{bmatrix}_{j}\right) = \nonumber\\
   \begin{bmatrix}
          x_i+v\Delta t \\
          ct_{i+1}
   \end{bmatrix}_{k}-
   \begin{bmatrix}
          x_i \\
          ct_i
   \end{bmatrix}_{k}
  \end{eqnarray}
or with the help of linearity of $T\left(v\right)$, simply:
  \begin{equation} \label{V}
  T\left(v\right)
   \begin{bmatrix}
         0 \\
         c\Delta t
   \end{bmatrix}_{j}=
   \begin{bmatrix}
         v\Delta t \\
         c\Delta t
   \end{bmatrix}_{k}
  \end{equation}
in which $\Delta t=t_{i+1}-t_{i}$.

This condition will be intuitive if we use the following definition of speed:
  \newtheorem*{Def}{Definition}
	\begin{Def}
	The speed of two events in some frame is defined to be the ratio of difference of their position coordinates to their time coordinates.
	\end{Def}
The definition may look peculiar, since we used the speed for two events instead of one. This is so because we cannot assign the term `speed' to a single event. In this fashion, a `moving point' consists of a very dense sequence of events such that the time interval and the spatial separation between every two event tend to zero keeping their ratio finite.
	\item The converse of $\left(3\right)$ also might be stated as:
	\begin{equation} \label{MINUSV}
  T\left(-v\right)
   \begin{bmatrix}
         0 \\
         c\Delta t
   \end{bmatrix}_{k}=
   \begin{bmatrix}
         -v\Delta t \\
         c\Delta t
   \end{bmatrix}_{j}
  \end{equation}
\end{enumerate}
\section{The speed of light}
As stated by the second axiom of SR, our transformation must also conserve the speed of light. The case is very similar to those of Eqs.~\eqref{V} and \eqref{MINUSV}. Thus we need events which their separation is defined by the speed of light, that is:
  \begin{eqnarray}
	 T\left(v\right)\left(
	 \begin{bmatrix}
         x_{i}+c\Delta t \\
         ct_{i+1}
   \end{bmatrix}_{j}-
   \begin{bmatrix}
          x_i \\
          ct_i
   \end{bmatrix}_{j}\right) = \nonumber\\
   \begin{bmatrix}
          x_i+c\Delta t \\
          ct_{i+1}
   \end{bmatrix}_{k}-
   \begin{bmatrix}
          x_i \\
          ct_i
   \end{bmatrix}_{k}
  \end{eqnarray}
which leads to:
  \begin{equation} \label{SL}
  T\left(v\right)
   \begin{bmatrix}
         c\Delta t \\
         c\Delta t
   \end{bmatrix}_{j}=
   \begin{bmatrix}
         c\Delta t \\
         c\Delta t
   \end{bmatrix}_{k}
  \end{equation}
It is important to note that this condition is directly related to the axioms of the theory, not to our physical intuition. We have enough empirical support to relate the unit of time to the unit of length using the speed of light as a universal constant. However, any other constant could do the job granted that it is independent of the reference frame.
\section{Lorentz transformation}
With axioms and conditions stated in the previous sections we are now at the stage to draw the transformation required. First of all, from Eqs.~\eqref{T} and \eqref{INV}:
\begin{eqnarray}\label{TWELV}
T^{-1}\left(v\right)&=& \frac{1}{\det T\left(v\right)}
  \begin{bmatrix}
         T_{22}\left(v\right) & -T_{12}\left(v\right) \\
         -T_{21}\left(v\right) & T_{11}\left(v\right)
  \end{bmatrix}\nonumber\\
  &=&T\left(-v\right)
\end{eqnarray}
Substituting Eq.~\eqref{TWELV} in Eq.~\eqref{MINUSV} then leads to:
\begin{equation} \label{SIXTEEN}
T_{12}\left(v\right)=vT_{11}\left(v\right)/c
\end{equation}

Now, from Eq.~\eqref{V} we have:
\begin{equation}
T_{12}\left(v\right)=vT_{22}\left(v\right)/c
\end{equation}
that is:
\begin{equation} \label{SEVENTEEN}
T_{11}\left(v\right)=T_{22}\left(v\right)
\end{equation}
In addition, from Eq.~\eqref{SL} we have:
\begin{equation} \label{EIGHTTEEN}
T_{12}\left(v\right)=T_{21}\left(v\right)
\end{equation}
Using Eqs.~\eqref{SEVENTEEN} and \eqref{EIGHTTEEN} one may simply write $T\left( v \right) = {\left[ {\det T\left( v \right)} \right]^2}{T^{ - 1}}\left( { - v} \right)$ which employing Eq.~\eqref{INV} gives
	\begin{equation}
	\det T\left( v \right) =  \pm 1
	\end{equation}
Noting that the mapping has to be continuous with respect to its parameter $v$ one is left with~\cite{SB}
	\begin{equation} \label{DET}
	\det T\left(v\right)=1
	\end{equation}
Equation.~\eqref{SEVENTEEN} together with Eqs.~\eqref{EIGHTTEEN} and \eqref{DET} gives:
\begin{equation} \label{NINETEEN}
T^{2}_{11}\left(v\right)-T^{2}_{12}\left(v\right)=1
\end{equation}
Squaring Eq.~\eqref{SIXTEEN} and substituting it into Eq.~\eqref{NINETEEN} gives:
\begin{equation}
T_{11}\left(v\right)=1/\sqrt{1-\left(v/c\right)^{2}}
\end{equation}

Using the following definitions
\begin{eqnarray}
 \beta&\equiv&v/c\\
 \gamma&\equiv&1/\sqrt{1- \beta^{2}}
\end{eqnarray}
one can write the transformation matrix in Eq.~\eqref{T} as:
\begin{equation} \label{TF}
 T\left(v\right)=
  \begin{bmatrix}
         \gamma & \beta\gamma \\
         \beta\gamma & \gamma
  \end{bmatrix}
\end{equation}
which is the well-known LT.

\section{Conclusion}
I have obtained the LT using a set of physically intuitive assumptions and the basic axioms of the SR. I have also shown that, on these bases, the transformation obtained is unique. The method presented in this paper can emboss the role played by the light speed as a universal constant in the theory. Despite all struggles with its concepts, I have shown taking the presumptions as true, guarantees the integrity of the SR and LT.

\section*{Acknowledgments}
The author sincerely appreciates the anonymous referee for having brought his attention to the paper by A.~Sfarti in addition to his patient perusal and helpful comments.

\end{document}